\documentclass{aa}  
\usepackage{graphicx}

\usepackage{epsf}
\usepackage{helvet}
\usepackage{xspace}
\usepackage{psfig}
\usepackage{longtable}
\usepackage{natbib}
\usepackage{txfonts}
\begin{document}
   \title{Sub-milliarcsecond precision spectro-astrometry of Be stars}

   \author{Ren\'e D. Oudmaijer \inst{1}  \and 
           A.M. Parr \inst{1} \and D. Baines \inst{2,1} \and J.M. Porter \inst{3}\thanks{deceased}  }

   \offprints{R.D.Oudmaijer}

   \institute{School of Physics and Astronomy, University of Leeds, Leeds LS2 9JT, UK\\
              \email{roud@ast.leeds.ac.uk}
         \and
National Physical Laboratory, Hampton Road, Teddington, TW11 0LW, UK 
\and
             Astrophysics Research Institute, Liverpool John Moores University, Twelve Quays House, Egerton Wharf, Birkenhead, CH41 1 LD, UK
             }

   \date{Received .. ; accepted ..}

 
  \abstract 
{The origin of the disks around Be stars is still not known. Further
  progress requires a proper parametrization of their structure, both
  spatially and spectrally. This is challenging as the disks are very
  small. }
{Here we assess whether a novel method is capable  of providing these data.  } 
{ We obtained spectro-astrometry around the Pa$\beta$ line of two
  bright Be stars, $\alpha$ Col and $\zeta$ Tau, to search for disk
  signatures. The data, with a pixel-to-pixel precision of the
  centroid position of 0.3..0.4 milliarcsecond is the most accurate
  such data to date. Artefacts at the 0.85 mas are present in the
  data, but these are readily identified as they were non-repeatable
  in our redundant datasets. This does illustrate the need of taking
  multiple data to avoid spurious detections. }
{  The data are compared with model simulations of the
  spectro-astrometric signatures due to rotating disks around Be
  stars. The upper limits we find for the disk radii correspond to
  disk sizes of a few dozen stellar radii if they rotate Keplerian.
  This is very close to observationally measured and theoretically
  expected disk sizes, and this paper therefore demonstrates that
  spectro-astrometry, of which we present the first such attempt, has
  the potential to resolve the disks around Be stars. }
{}

   \keywords{techniques: high angular resolution - techniques:
   spectroscopic - stars: individual: $\alpha$ Col, $\zeta$ Tau -
   stars: emission-line,Be } 

   \maketitle

\section{ Introduction }

For decades it had been surmised that Be stars are surrounded by
disk-like structures. At first this was inferred by indirect means
such as the doubly peaked H$\alpha$ emission line profiles
\citep{struve_1931} and polarization \citep{Poeckert:1975}. This
notion was confirmed only much later by direct, interferometric
observations at radio wavelengths \citep{dougherty_1992}. Later,
dedicated long-baseline optical and near-infrared (NIR) interferometry
resolved the disks at selected baselines
(e.g. \citealt{quirrenbach_1997, tycner_2004, meilland_2007} - for a
general review on Be stars see \citealt{porter_review}).  So far,
relatively few Be stars have been studied in this manner.  This is due
to the fact that the disks are small, even the largest observed disks
are typically of order a few milliarcsec (mas) in diameter and
observations remain challenging.

In this paper we investigate the potential of spectro-astrometry to
detect disks around Be stars. This technique is a powerful tool; it
enables us to investigate small scale structures with a standard
instrumental set-up.  In addition, since data are taken at high
spectral resolution, it also allows kinematical studies to be
performed at superior resolution than for interferometry.
Spectro-astrometry is a proven method to study otherwise unresolved
structures in longslit spectra. It has been used to study binaries
(\citealt{bailey_1998,baines_2004,baines_2006, schnerr_2006}),
outflows from young objects \citep{takami_2003}, disks around young
objects \citep{ponto_2008} and even made possible the discovery of
bi-polar jets from Brown Dwarfs \citep{whelan_2005}.  Conceptually,
this technique is straightforward, it measures the relative spatial
position of spectral features from a longslit spectrum. For example,
the red- and blueshifted emission of a rotating disk will be located
on opposite sides of the continuum. Even when spatially unresolved,
the centroid position of the spectrum will be offset from the
continuum, and this can be determined very accurately to sub-pixel
values (see e.g. \citealt{bailey_1998}).  The method has been shown to
detect binaries at separations of 0.1 arcsec in conditions where the
seeing was in excess of 2 arcsec, while brightness differences between
the binary components of up to 6 magnitudes have been observed as well
\citep{baines_2006}.  Observationally, it is a comparatively cheap
method, requiring only a stable spectrograph and a digital detector.
It can therefore be applied to large samples of object.

As for example demonstrated by \citet{takami_2003}, the positional
accuracy of the centroid mainly depends on the number of photons and
the seeing and can be expressed as $ \sigma = 0.5 \times {\rm FWHM}
\times N^{-\frac{1}{2}} $, with the error $\sigma$ and full width half
maximum of the profile (typically the seeing) expressed in arcsec or
milliarcsec, and $N$ is the number of photons.  For shot-noise
dominated statistics, $N^{-\frac{1}{2}}$ equates to the inverse
signal-to-noise ratio (SNR) of the total spectrum.  Therefore, the
requirements are proper sampling, high SNR, and a narrow instrumental
point spread function (i.e. good seeing). \citet{baines_2006}
achieved a position accuracy, as measured from the root-mean-square
(rms) variations in the position, of 2 mas in 2 arcsec seeing.
The aim of the present study is to significantly improve on this
statistic to investigate whether we can detect the milliarcsecond
scale disks around Be stars.

 This paper is organized as follows. In Sec. 2 we describe the
observations of two bright, nearby Be stars.  and reduction
procedure. In Sec. 3 we present the results of the study, introduce
the sub-milliarcsecond spectro-astrometry and we discuss the results
in terms of a simple model. We conclude in Sec. 4.

\begin{table*}
\begin{center}
\begin{tabular}{@{}lllllrllllll}                                
                                
  \hline
  Object      & HD/HR  & Sp type &{\it V}& vsin$i$  &{PA}    & exp time & SNR & Pa$\beta$ EW &  $\Delta \, v$ & rms pos\\
              &        &         & (mag) &(kms$^{-1})$  &($^{\rm o}$)&  (s)   &     &   ({$\rm \AA$})&   (kms$^{-1}$)  & (mas)   \\
\hline                                
 $\alpha$ Col & 37795  & B7IVe   & 2.64  &  176     &  90-270 & 8 x 6   & 1200 & $-$8        &  134           & 0.35 \\
              &  1956  &         &       &          & 180-360 & 8 x 6   &      &             &                & 0.40 \\
$\zeta$ Tau   & 37202  & B4IIIpe & 3     &  310     &  58-238 & 8 x 10  & 1500 & $-$5        & 226            & 0.25 \\
              & 1910   &         &       &          & 148-328 & 8 x 10  &      &             &                & 0.30 \\
\hline
\label{table}
\end{tabular}
\caption{The targets and observational details. The fundamental
parameters are taken from the Bright Star Catalog \citep{bsc}. $\Delta \,
v$ denotes the peak separation of the line profiles. The SNR is
measured in a line free region of the total, co-added, spectra. The
rms in the positions are also measured in line free regions, this time
in the co-added data in each orientation. }
\end{center}                                                  
\end{table*}

\section{Observations and Data Reduction}

For this experiment we selected two Be stars that were bright,
close-by and had a track record of strong hydrogen recombination line
emission. These factors should ensure that they are surrounded by
comparatively large disks. Indeed, $\zeta$ Tau had been measured to
have an H$\alpha$ diameter of 7 mas
(e.g. \citealt{tycner_2004}). $\alpha$ Col has no published
interferometric data, estimates indicate a larger size of its line
emitting region than of $\zeta$ Tau \citep{dachs_1992}.  For the
choice of telescope, we had to trade-off between an excellent sampling
of the spectro-astrometry and the choice of target line. H$\alpha$ may
be expected to form in larger regions, but the near-infrared
instrumentation described below provided excellent sampling. For this
pilot study, we decided to push the best available equipment to us and
observed Pa$\beta$ at 1.28$\mu$m.

The spectro-astrometric data were obtained in service mode with the
Phoenix instrument \citep{hinklephoenix} mounted on 8m Gemini South in
Chile during the night of December 19 (UT) 2004.  Phoenix is a high
resolution near-infrared spectrometer operating in the wavelength
region 1-5$\mu$m.  The target line was the Pa$\beta$ hydrogen
recombination line, a strong line in a region of the spectrum that is
relatively unaffected by telluric absorption.

The grating was set such that the 1.28 $\mu$m Pa$\beta$ line was in
the centre. The detector was an Aladdin 1024$\times$1024 InSb array
with a pixel size of 5.9$\times10^{-6} \, \mu$m (in wavelength,
corresponding to 1.4 kms$^{-1}$) spectrally, resulting in an unvignetted
wavelength coverage of 1300 kms$^{-1}$.  The pixel size in the spatial
direction was 85 milliarcsec, and the slit was 4 pixels wide. The
seeing during the observations was hovering between 0.40-0.55 arcsec
as measured from the central parts of the spectra. 
The set-up
ensured that both the spatial and spectral resolution elements were
sampled by 4-5 pixels. This is much better than Nyquist sampling and
having the data sampled by such a large number of pixels is a crucial
constraint when dealing with spectro-astrometry. Although not strictly
necessary with such bright targets at this wavelength, the
observations were done in the standard a-b-b-a nodding on the slit to
remove sky emission.

In addition to the usual East-West and North-South slit positions,
that makes sure we can measure the position angles for any extended
material in the data, we observed at the opposite angles as well.  The
rationale is that the observations have to be repeatable and spurious
effects should be identified by multiple observations. Real effects
are each other's mirror image (one is looking at the objects upside
down as it were), instrumental artifacts would be present in the same
direction on the array (see for more details \citealt{bailey_1998} and
\citealt{baines_2006}).  In the case of $\alpha$ Col, the slit
position angles (PA) were set at 0$^{\rm o}$, 90$^{\rm o}$, 180$^{\rm
o}$, and 270$^{\rm o}$. $\zeta$ Tau was observed at 58$^{\rm o}$,
148$^{\rm o}$, 238$^{\rm o}$, and 328$^{\rm o}$. The choice of
different PAs rather than the usual EW-NW settings was that the slit
would be aligned with the disk resolved in interferometric data of
\citep{quirrenbach_1997}. However, they report a position angle of
$-58^{\rm o}$, and the omission of the minus sign in our instrumental
set-up means that the data are somewhat less efficient than they could
have been.  At each slit position, the objects were observed 4 times
with the nodding procedure and the total spectra consist of 16
integrations with exposure times of 6s each for $\alpha$ Col and 10s
for $\zeta$ Tau respectively. The overhead associated with rotating
the slit was small and the total on-target time was
less than 45 minutes in both cases.

\begin{figure*}
\begin{center}
\includegraphics[width=0.95\textwidth]{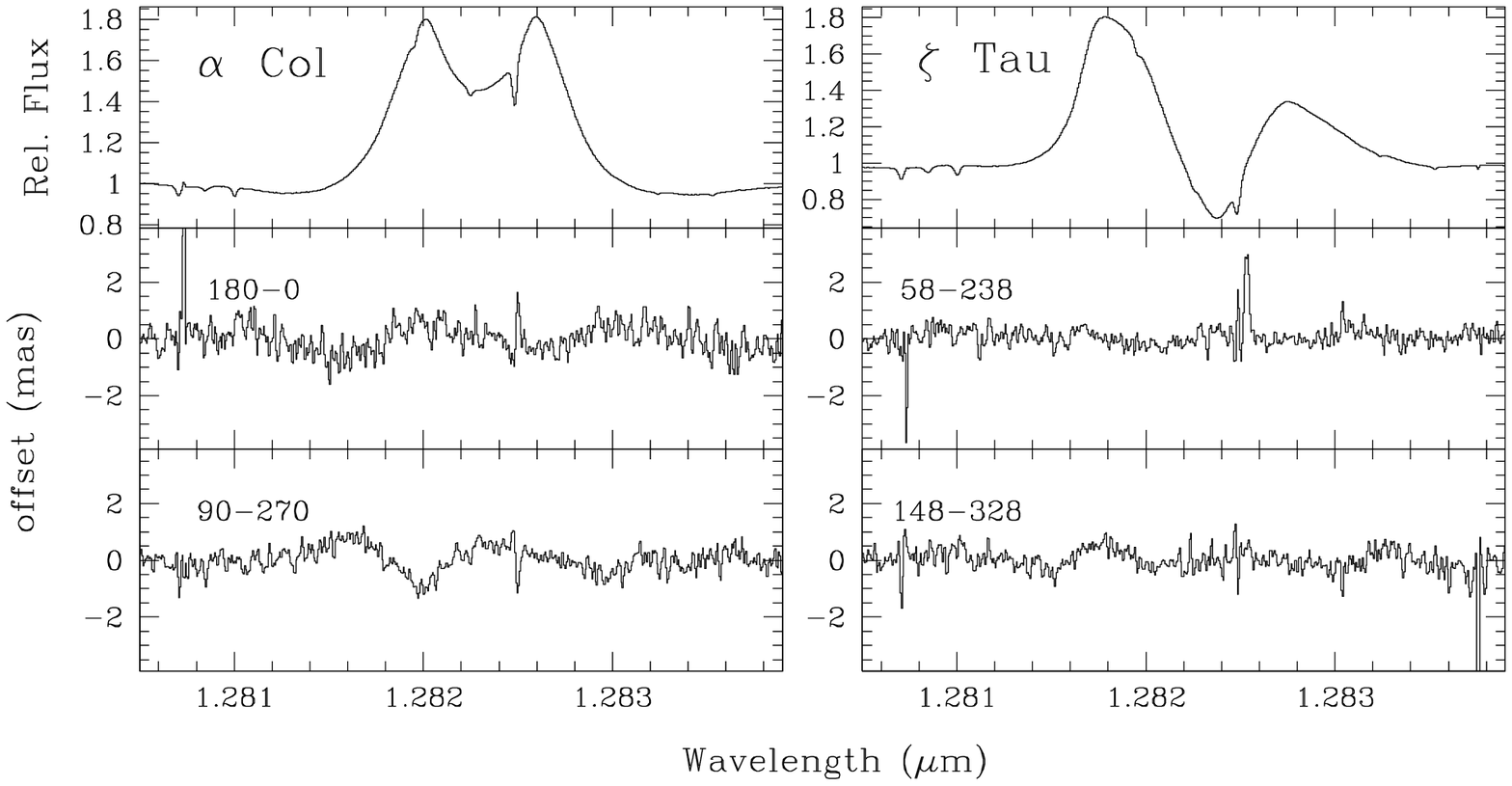}
    \caption{Data on $\alpha$ Col (right) and $\zeta$ Tau (right). 
The top panels represent the total intensity spectra. The bottom
panels are the photo-centers expressed in mas, at perpendicular
orientations whose values are indicated in the figures.
 \label{specast}}
\end{center}
\end{figure*}

The data were reduced in a standard manner for optical data using both
the IRAF \citep{iraf} and Starlink software packages. Dark frames were
subtracted from the original frames, which were then divided by a
normalized flatfield. The intensity spectra were extracted and the
individual spectra co-added to arrive at the total spectra that are
discussed in the remainder of this paper.  The observational details
and some derived parameters are summarised in Table 1 The wavelength
calibration was performed by identifying telluric absorption lines
measured from the catalogue by \citet{hinkle_1995} and finding the
dispersion of the spectra. The resulting wavelength should be accurate
to within 0.25 kms$^{-1}$.  The FWHM of telluric lines that were
assumed to be unresolved was measured to be in the range of 6..6.5
kms$^{-1}$, in agreement with the expected resolution $\lambda /
\Delta \lambda \sim 50,000$.

Spectro-astrometric information was extracted from the 2-dimensional
longslit data by fitting a Gaussian profile to the stellar flux at
each pixel in the spatial direction.  By visually inspecting the data
we confirmed that the Gaussians are a good representation of the point
spread function. However, as \citet{porter_2004} pointed out, the
precise shape of the fitting function is not very critical.  The
positions of the center were recorded as a function of pixel number,
and later put on a wavelength scale.

Four position spectra were obtained at every position angle. The
individual traces had different shapes on the larger (more than tens
of pixels) scales. This effect is presumably present because they were
recorded at different locations on the array because of the nodding
procedure. To eliminate these large scale fluctuations, while at the
same time preserving the smaller scale properties, the position
spectra were fitted by a high order polynomial.  The position data
taken at opposite position angles (0-180$^{\rm o}$ etc), were
subtracted from each other to minimize any remaining
instrumental effects. Prior to combining, all eight traces per
orientation were visually inspected in order to identify and remove outlying
data. For both orientations of $\alpha$ Col, one trace each was
discarded, for $\zeta$ Tau all traces at a PA of 58$^{\rm o}$ were
retained, while 2 of the 8 traces perpendicular to this were
excluded. 

The precision in determining the photo-centre of the longslit
spectra as measured from pixel-to-pixel deviations from the mean
position was of order 0.3 mas (see Table 1).  This positional accuracy
is the best quality data ever published from a statistical
pixel-to-pixel point of view. However, in the current data set we also
see variations, unrelated to the pixel-to-pixel variations, over
larger scales, with excursions up to 1 mas.  The largest such
multi-pixel variations can be seen in the East-West direction for
$\alpha$ Col (see Figure~\ref{specast}). A sine-wave type feature
spanning many pixels with an amplitude of 0.85 mas, almost 3$\sigma$,
appears visible. The dip corresponds to the blue peak of the Pa$\beta$
line profile, but the two local ``maxima'' around the dip do not
correspond to any obvious feature in the emission line.  Further
inspection of the data revealed that the individual traces taken at
opposite angles also show this pattern.  Significantly however, the
minima and maxima occur at different wavelengths, and are therefore
not reproduceable.  A similar behaviour is also found in the data of
$\zeta$ Tau, but in this case it is cancelled out after combination of
the individual traces. We therefore conclude that the traces in
$\alpha$ Col that seem to display a ``periodic'' signal with an
amplitude of less than a hundredth of a pixel are artefacts, as they
are not reproduceable in our, redundant, data. The reasons for this
are unclear and future observations are planned to investigate this
issue. In summary, the data have a, statistical, precision of 0.3 mas,
while larger scale artefacts of order slightly less than 1 mas are
identified.

\section{Results }

\subsection{The Pa$\beta$ lines}

The results for $\alpha$ Col and $\zeta$ Tau are plotted in
Fig.~\ref{specast}. The top panels display the total intensity spectra
while the bottom two panels represent the spectro-astrometry at the
two orientations respectively. Let us first discuss the Pa$\beta$
profiles. Both stars have doubly peaked emission lines.  $\alpha$ Col
has a regular, symmetric line profile, while the blue peak of $\zeta$
Tau's emission is much stronger than the red peak.  The Equivalent
Widths ($W_{\lambda}$) are $-$8 and $-$5~$\rm \AA$ and the line peak
separations are 134 and 226 kms$^{-1}$ for $\alpha$ Col and $\zeta$
Tau respectively.

The emission at the wavelengths covered by the Pa$\beta$ line is due
to three components. Firstly, the emission line itself, secondly
continuum free-free emission and thirdly, the stellar continuum, which
is diluted by the underlying, photospheric, Pa$\beta$ absorption line.
We can estimate the contribution of the free-free emission to the
total flux. At shorter wavelengths it is fairly low, as the continuum
excess due to free-free emission increases towards longer wavelengths
(see e.g. the in-depth study by \citealt{dougherty_1991}).  For
$\zeta$ Tau, \citet{dougherty_1991} derive an excess of 0.1 magnitude
at 1.25 $\mu$m, which roughly corresponds to 10\% of the emission
being due to the disk. \citet{dougherty_1991} did not include $\alpha$
Col in their sample.  \citet{dachs_1988} observed the object in the
optical and near-infrared one month apart and we derive a
quasi-simultaneous {\it V$-$J} colour of $-0.18$. According to
\citet{koornneef_1983}, {\it (V$-$J$)_{0}$} for a B7V object is
$-$0.25 mag (he does not list values for sub-giants with luminosity
class IV). Taken at face value, we would thus obtain a non-physical,
negative excess. However, the difference is comparable to the
uncertainty in spectral class and photometric errorbars, and
illustrates that the excess continuum emission at Pa$\beta$ must be
very small.  The depth of the underlying photospheric Pa$\beta$
absorption can be assessed using the data of
\citet{wallace_2000}. They present medium resolution spectroscopy of
88 MK spectral standards, amongst which a number of B-type stars. The
stars in their sample with spectral types closest to ours, B7III, B7V
and B3IV (HR 1791, HR 3982 and HR 6588 respectively), span a wide
range in spectral type. The central dip ranges from 0.6..0.65 of the
continuum for the narrower lines (B3IV, B7III) to 0.78 (B7V) for the
broader line. At $-$100 and +100 kms$^{-1}$ from the line center,
where the line emission peaks are found, the absorption for all three
objects reaches down to 0.85..0.9 of the continuum, i.e. a depression
of 10..15\%.

Hence, given that the spectral types of our target objects of B4III
and B7IV, are similar to that sampled by these MK standard stars, we
assume that the photospheric absorption line underneath the line peaks
is about the same fraction of the line free continuum.  As the peak
line emission is roughly twice that of the stellar continuum (1.98 and
1.84 for $\alpha$ Col and $\zeta$ Tau respectively), we find that the
emission from the stellar photosphere and the hydrogen recombination
line are approximately equal.

\subsection{The spectro-astrometry} 

The spectro-astrometric traces, at both orthogonal orientations, are
shown in the middle and bottom panels of Fig.~\ref{specast}.  The
traces are normalized to the stellar continuum and the deviations from
it are expressed in milliarcseconds. Several things are immediately
apparent from the data. Firstly, the rms variations around the mean
position are much smaller than 0.5 mas (Table 1). Thus, the data have
a sub-milliarcsecond precision, and therefore constitute the most
accurate spectro-astrometry of any object hitherto observed. As we are
exploring unknown territory, it may not come as a surprise that we
encounter new problems in the data. After the multi-pixel variations
that stretch to slightly less than 1 mas in $\alpha$ Col, the second
obvious finding in the spectro-astrometric traces is that the telluric
features show strong, in fact the strongest, signals. These narrow
absorption lines are unresolved and we suspect that pixellation
effects give rise to these features. Checks on data taken at opposite
angles revealed that the amplitude of the excursion varies with the
location of the longslit spectrum on the array. It is always in the
same direction, and therefore an artefact. For illustration we kept
the data in the figures. Finally, there is no obvious excursion in
position space associated with the Pa$\beta$ profiles.

We conclude that in the present data we find no evidence for
significant features in the positional data down to sub-milliarcsecond
levels. It also illustrates the need for multiple exposures to ensure
consistency and to avoid artifacts in the data being interpreted as
real. In the following we will investigate the implications for the
presence of disks around both objects and the presence of a binary
companions.

\subsection{On the stars' binarity and their disks' sizes}

Based on radial measurements $\zeta$ Tau is reported to be a
single-lined, close binary \citep{jarad_1987}. The separation is 5 mas
and, from the mass function, the primary is at least 5 magnitudes
brighter than the secondary \citep{tycner_2004}. This large
magnitude difference combined with the small separation makes
detection of the secondary very difficult (see also
\citealt{baines_2006}), and explains why we do not see a binary
signature in the data of $\zeta$ Tau. $\alpha$ Col has not been
reported to be a binary, and the  data do not show the presence
of a binary companion either.

We can estimate the size of the Pa$\beta$ line emitting region from
the spectrum.  Most methods such as the $W_{\lambda}$ of the emission
employ the entire line profile to do this
(e.g. \citealt{grundstrom_2006}), even interferometrically determined
sizes are based on the total line emission. Here, we exploit the fact
that we have spatial information available at high spectral
resolution.  If the disks are in Keplerian rotation, we can compute
the distance from the star of the bulk of the orbiting material using
its rotation velocity. Using values for the masses and radii for the
spectral types (taken from \citealt{str_kur}, and interpolated between
B3 and B5 to arrive at a value for $\zeta$ Tau), we computed the
Keplerian rotation speeds at the stellar surface (457 and 489
kms$^{-1}$ for $\alpha$ Col and $\zeta$ Tau).  The observed velocities
at the line peaks (half the peak separation in Table 1) combined with
the distances to the objects provided by Hipparcos, yield the distance
of the line peak forming regions from the star of 8.8 mas ($\alpha$
Col) and 3.9 mas ($\zeta$ Tau).  However, the observed velocities are
smaller than the true value by a factor sin$i$, and the distance from
which the emission originates is smaller by (sin$i$)$^{2}$ (for
Keplerian rotation).  Taking the inclinations derived by
\citet[45$^{\rm o}$ and 66$^{\rm o}$]{fremat_2005} we obtain 4.4 and
3.2 mas for $\alpha$ Col and $\zeta$ Tau respectively.  The amplitude
of the excursion in the positional data associated with these
separations can be calculated by simulating the data at the line-peak,
convolving them with the seeing and determining the
spectro-astrometric trace (cf. \citealt{baines_2006}). In both cases,
the photo-center is precisely halfway because the line peaks are
equally bright as the star.  We therefore would expect, based on the
above, that in the present data line excursions of up to 2.2 mas
($\alpha$ Col) and 1.6 mas ($\zeta$ Tau) can be observed.  This is the
maximum observable separation, as the above computation assumes that
all line emission arises from a thin ring with a rotation speed
corresponding to the line peak.

\begin{figure}
\begin{center}
\includegraphics[width=0.5\textwidth]{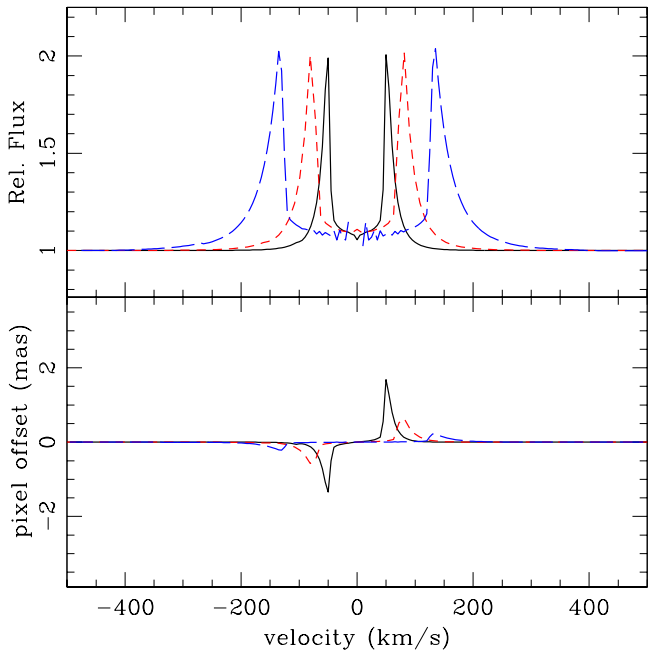}
    \caption{Toy model predictions of the spectro-astrometric signature
    of a disk surrounding a Be star with properties similar to the
    target objects. The top panel shows the flux spectra, the bottom
    panel shows the predicted astrometry for the set-up used in our
    observations for three different cases, an outer radius of 70
    stellar radii (solid line), 30 stellar radii (short-dashed), 10
    stellar radii (long-dashed).
 \label{specastmod}}
\end{center}
\end{figure}

Not all emission at the line peak comes from a single ring however, as
the projected velocities for smaller, faster, rings will be observed
as well at the observed Doppler shifts.  To assess this effect we
performed some simple model calculations. We assume the star to be
surrounded by a geometrically thin, Keplerian rotating disk reaching
onto the stellar surface, with the line flux per unit area following a
simple power law in radius.  The main input parameters of the model
are the stellar radius, rotation speed, the inclination and emission
line strength (which are all fairly well known), the remaining free
parameters are the disk's outer radius and the exponent of the power
law. The model produces a two dimensional position-velocity diagram,
which is binned up and smoothed to represent our pixel sizes of
5~kms$^{-1}$, 85 mas and seeing of 500 mas respectively.  From the
resulting data, the spectro-astrometry is then measured.  Changing the
outer radius of the model disk increases the spectro-astrometric
excursions, which then occur at lower velocities, as expected from
Keplerian rotation. A stronger line flux will yield a larger
spectro-astrometric excursion because the photo-centre shifts more in
the direction of the emission line.

In the extreme case of optically thick emission, the powerlaw will
have a flat slope and the emerging line flux will be dominated by the
outer parts of the disk.  In the other extreme, that of optically thin
emission, the power law depends on the density distribution. For an
isothermal, flaring Keplerian disk, the surface density, and by
implication the flux per unit area, has an $r^{-2}$ powerlaw
dependence (cf. \citealt{carciofi_2006}). As a consequence, the line
emission moves towards the inner parts of the disk. The main
positional excursions will thus occur at higher velocities, closer to
the star and therefore be smaller than in the optically thick case.
Changing the exponent of the powerlaw also affects the shape of the
emission line. A shallower exponent, more representative of the
optically thick case, puts more flux at lower velocities, while a
steeper power law, closer to the optically thin situation, results in
narrower lines, with the line peak at higher velocities.

In general though, unless the exponent gets too steep, the excursions
are of similar magnitude when the same line-to-continuum ratio is
simulated.  We performed a large parameter study, but for the purposes
of this paper, we will restrain ourselves to one illustrative example
representative of both objects.  We set the line-to-continuum ratio to
be 2 (as per the spectra in Fig.~\ref{specast} and derived above), use
a stellar rotational velocity of 475 kms$^{-1}$ (halfway the values
for both objects) and an inclination of 55$^{\rm o}$ (also roughly
halfway the two objects) and use a stellar radius of 0.2 mas. For the
outer radius of the disk we take 10, 30 and a maximum of 70 stellar
radii (cf. \citealt{marlborough_1997}).  The resulting data are shown
in Fig.~\ref{specastmod}.  The top panel presents the resulting model
line profiles.  As expected, the lines are doubly peaked with peak
separations that are larger for smaller disk radii. The separations
range from $\sim$110kms$^{-1}$  for the largest disk to 160 and 270 kms$^{-1}$
for the smallest disks, respectively. This trend is explained by the
fact that these velocities correspond to the Keplerian rotation speeds
at the maximum possible radii, where most of the line flux originates
if the emission is optically thick.  The most notable differences
between the model line profiles and the observed line profiles are the
relative narrowness of the line peaks and the little emission at low
projected velocities. This is probably due to the fact that the model
disks are assumed to be geometrically thin, resulting in low projected
emitting surface areas at low velocities. In reality the disks are
flared, and therefore the emitting area will be much larger, in
particular at these low velocities. In addition, line broadening is
not taken into account here.  For a proper treatment, radiative
transfer models such as those by \citet{carciofibjorkman_2006} will be
an excellent tool. Using such advanced models is beyond the scope of
this paper, in which we wish to obtain a rough figure for the excursions
only.

We also note that the blue peak of $\zeta$ Tau is much stronger than
the red peak.  This is most likely due to one-armed oscillations in
its disk, which give rise to such asymmetry, as for example detected
for $\zeta$ Tau at the 0.7 mas level by
\citet{vakili_1998}. Accordingly, the positional excursion of the blue
peak would be larger than that of the red one, but its signature would
not affect the overall appearance of the spectro-astrometry.

Moving to the spectro-astrometric signature in the simulations, we
find that the positional excursions are smaller for smaller disks
(1.5, 0.6 and 0.2 mas respectively) for the same reason as that the
peak separations are larger: most flux comes from the outer parts, and
the largest model disk will naturally result in the largest detection.
As the orientations of the disks are not aligned with the slit
positions, we might in reality observe smaller excursions in our data
by up to a factor of 0.7 due to projection effects.

The bottomline of the exploratory model simulations is that disks with
a size of order 70 stellar radii and a predicted excursion of 1.5 mas
could just about have been observed at the 3-4 $\sigma$ level, while
the 30 stellar radii disk would have been a 2 $\sigma$
detection. These limits are approaching the real sizes of the disks.
According to \citet{tycner_2004}, the disk of $\zeta$ Tau has a
diameter of 7 mas, and thus a radius of $\sim$18 stellar radii,
whereas $\alpha$ Col's disk is assumed to be larger, mainly because of
its larger line EW \citep{dachs_1992}. It is clear that our
pixel-to-pixel precision, of order 0.35 mas, is reaching that needed to
detect Keplerian disks.  It will thus be possible to measure the disks
and their kinematics and therefore to properly constrain the disks
with future data. In order to achieve this, observations should 
result in a better SNR, or be targetted at stronger emission lines (in terms
of line-to-continuum ratio), either for stars with larger disks or
from intrinsically stronger lines at different wavelengths such as
H$\alpha$.

High precision data such as these combined with the latest radiative
transfer models \citet{carciofibjorkman_2006} will allow us to be in a
position to fully constrain the kinematical structure of Be star
disks, and reveal their origin.


\section{Conclusion}

In conclusion, we employed high precision spectro-astrometry to assess
the potential of the method to detect the disks around two Be
stars. We achieved rms variations in the position spectra of order 0.3
mas, the highest precision spectro-astrometric data in the literature.
We did not detect any features related to the Pa$\beta$ lines, but
found artefacts at the 1 mas level. These were easily identified as
they were non-repeatable in our redundant datasets.

Simple, robust, estimates of the size of the line forming regions
showed that the current set-up was on the limit of detecting the
disks, if they were rotating Keplerian. Indeed, the method has the
potential to distinguish between Keplerian rotating disks and angular
momentum conserving disks which would be much smaller.  This study,
the first of its sort, has shown that the method has great potential
in probing small, sub-milliarcsecond, scale structures and future
observations using improved set-up, even higher SNR, and possibly
moving to the intrinsically brightest hydrogen recombination lines which
should reveal even larger disk-signatures in the data are planned.

\subsection*{Acknowledgments}

RDO is grateful for the support from the Leverhulme Trust for awarding
a Research Fellowship.  AMP acknowledges support from The Rothschild
Community of Excellence Programme. This work is based on data from the
Phoenix infrared spectrograph, developed and operated by the National
Optical Astronomy Observatory. The observations are from programme
GS-2004B-Q-92 obtained at the Gemini Observatory, which is operated by
the Association of Universities for Research in Astronomy, Inc., under
a cooperative agreement with the NSF on behalf of the Gemini
partnership.

\bibliographystyle{aa}
\bibliography{mnemonic,RenesRefs}

\label{last page}
\end{document}